\documentclass[12pt,a4paper]{article}
\pdfoutput=1
\usepackage{amsthm,bm,amsmath,amssymb,epsfig,multicol,multirow}
\usepackage[square, numbers]{natbib}
\usepackage{graphics,rotating,graphicx}
\usepackage{array,multirow,caption2}
\usepackage{amssymb}
\usepackage{amsthm}
\usepackage{bm,amsmath,multicol}
\usepackage{booktabs}

\newcommand{\asim}{\stackrel{\text{\tiny approx.}}{\sim}}

\def\e{\hbox{E}}

\def\var{\hbox{Var}}

\begin{document}

\noindent{\Large \textbf{Meta-analysis of ratios of sample variances}}



\vspace{0.5cm}

\noindent LUKE A. PRENDERGAST

\noindent \textit{Department of Mathematics and Statistics, La Trobe
University, Melbourne, Australia, 3086}

\vspace{0.5cm}

\noindent ROBERT G. STAUDTE

\noindent \textit{Department of Mathematics and Statistics, La Trobe
University, Melbourne, Australia, 3086}
\vspace{1cm}

\noindent\textbf{ABSTRACT.  When conducting a meta-analysis of standardized mean differences (SMDs), it is common to assume equal variances in the two  arms of each study.  This leads to Cohen's $d$  estimates for which interpretation is simple.  However, this simplicity should not be used as a justification for the assumption of equal variances in situations where evidence may suggest that it is incorrect.  Until now, researchers have either used an $F$-test for each individual study as a justification for the equality of variances or perhaps even conveniently ignored such tools altogether.  In this paper we propose using a meta-analysis of $F$-test statistics to estimate the ratio of variances prior to the combination of SMD's. This procedure allows some studies to be included that might otherwise be omitted by individual fixed level tests for unequal variances, sometimes occur even when the assumption of equal
variances holds. The estimated ratio of variances, as well as associated confidence intervals, can be used as guidance as to whether the assumption of equal variances is violated.  The estimators considered include variance stabilization transformations (VST) of the $F$-test statistics as well as MLE estimators.  The VST approaches enable the use of QQ-plots to visually inspect for violations of equal variances while the MLE estimator easily allows for the introduction of a random effect.  When there is evidence of unequal variances, this work provides a means to formally justify the use of less common methods such as log ratio of means when studies are measured on a different scale.}

\vspace{0.5cm}



\section{Introduction}

 The $F$-distribution (see, for e.g., Chapters 27 and 30 of \cite{JO&KO&BA95}) is encountered in many applications of statistics including  the testing of equality of variances, analysis of variance (ANOVA) and more broadly a test for the overall model in linear regression analyses.  A problem that may be of interest to many is how best to combine the evidence from several independent $F$-distributed random variables to improve power.  For example, suppose that we have $K$ independent studies reporting $F$-statistics (or summary measures that can be used to calculate an $F$-statistic) for the testing of the same hypotheses.  Rather than relying on $K$ individual hypothesis tests, we seek to combine these $F$-statistics to obtain a single measure that can be used to test the hypotheses with increased power or to obtain an improved estimation of an associated parameter.

Of particular interest here is a meta-analysis for standardized mean differences (SMDs). In each study there are two  populations with means $\mu_1$ and $\mu_2$ and the same variance $\sigma^2$, and the population SMD is
$\delta = (\mu_1-\mu_2)/\sigma$. An estimate of $\delta$ is Cohen's $d$ \citep{CO88}, denoted $d=(\overline{x}_1-\overline{x}_2)/s_p$, where $\overline{x}_1$ and $\overline{x}_2$ are sample means estimates for $n_1$ and $n_2$ sampled observations from each of the populations and $s_p^2$ is the pooled sample variance estimate of $\sigma^2$.  Cohen \cite{CO88} suggested a `rule of the thumb' for the SMD where 0.2 is considered small, 0.5 medium and 0.8 large and values of $d$ from studies with data collected on different measurement scales can be readily compared.  It is important here to ensure that the assumption of equal variances is justified, or at least that there is a lack of evidence to suggest that the population variances are not equal, and one possibility is to conduct an $F$-test for equality of variances \citep[p.323]{JO&KO&BA95} in each study.  Our interest is on the availability of more than one study and on whether the assumption of equal variances is justified in general.

While many tests can be reasonably robust to even moderate departures from equal variances (e.g. ANOVA), this is an entirely different scenario.  Here, the equal variances assumption is used to obtain a simple-to-interpret measure, $\delta$, or its variations.  If the variances are not equal, then the estimate $s_p$ may not resemble anything like the variances in each arm and the resulting estimate $d$ may be misleading.  We emphasize this point in the next section.

In Section 2 we will provide a motivating example before discussing the transformation of $F$-statistics in Section 3.  Maximum likelihood estimators are discussed in Section 4 where both fixed effect and random effects models are considered.  Simulations are provided in Section 5 that assess the performance of estimators of the ratio of variances.  In Section 6 we reconsider the example in Section 2 and carry out the meta-analysis for $F$-statistics.  Concluding remarks are provided in Section 7.

\section{A motivating example}\label{section:mot}

 Thakkinstian {\em et al.} \cite{TH04} reported on 15 studies with the data shown for 13 of these given in Table \ref{table:TH04} below.  Since two of the 15 studies were not used in their meta-analysis of SMDs which we consider shortly, we have not included these studies in our table.

\setlength{\tabcolsep}{4pt}
\begin{table}[ht]
\centering
\begin{small}
\begin{tabular}{rrrrrrrrrrrrrrrrrr}
  \hline
  \multirow{2}{*}{Study} & \multicolumn{3}{c}{BB} &  & \multicolumn{3}{c}{Bb} & & \multicolumn{3}{c}{bb} & &  \multicolumn{3}{c}{Bb/bb}\\
  & $n$ & Mean & SD & & $n$ & Mean & SD & & $n$ & Mean & SD && $n$ & Mean & SD\\ \cmidrule{1-4} \cmidrule{6-8} \cmidrule{10-12} \cmidrule{14-16}
  1 &      7 &0.970 &0.160&&  35& 1.040& 0.170&&  34& 1.000& 0.190&&  69& 1.020& 0.180 \\
  2 &      2 &1.077 &0.011&&  14& 1.083& 0.099&&   7& 1.099& 0.171&&  21& 1.088& 0.123 \\
  3 &     15 &1.007 &0.158&&  36& 1.047& 0.227&&  40& 1.003& 0.166&&  76& 1.024& 0.197 \\
  4 &     12 &0.980 &0.150&&  19& 0.970& 0.120&&  18& 1.000& 0.130&&  37& 0.985& 0.124 \\
  5 &     38 &0.880 &0.160&& 134& 0.870& 0.110&&  96& 0.860& 0.130&& 230& 0.866& 0.119 \\
  6 &     77 &0.906 &0.153&& 276& 0.932& 0.136&& 196& 0.924& 0.128&& 472& 0.929& 0.133 \\
  7 &      8 &0.870 &0.090&&  43& 0.860& 0.160&&  52& 0.890& 0.150&&  95& 0.876& 0.155 \\
  8 &    107 &0.870 &0.180&& 306& 0.870& 0.160&& 175& 0.870& 0.150&& 481& 0.870& 0.156 \\
  9 &     71 &0.810 &0.253&& 219& 0.846& 0.186&& 120& 0.897& 0.136&& 339& 0.864& 0.172 \\
  10 &    46 &1.034 &0.177&&  98& 1.024& 0.137&&  56& 1.041& 0.122&& 154& 1.030& 0.132 \\
  11 &    27 &0.863 &0.152&&  72& 0.871& 0.167&&  62& 0.929& 0.124&& 134& 0.898& 0.151 \\
  12 &    25 &0.924 &0.145&&  34& 0.951& 0.138&&  21& 0.944& 0.131&&  55& 0.948& 0.134 \\
  13 &    19 &0.651 &0.078&&  59& 0.718& 0.070&&  24& 0.723& 0.083&&  83& 0.719& 0.074 \\
  \hline
\end{tabular}
\end{small}
\caption{Data from Table 2 of \cite{TH04} (excluding two studies that were not part of their meta-analysis).  Sample sizes ($n$), means and the standard deviations (SD) for three genotype groups and including the combined grouping of Bb and bb (Bb/Bb).}\label{table:TH04}
\end{table}

There are three groups for each of the 13 studies; namely BB, Bb and bb which refer to geno-types of individuals studied.  The estimate is the mean spinal Bone Mass Density (BMD) for premenopausal women and the question is whether there is a difference in mean BMD with respect to geno-type.   In a meta-analysis of SMDs, Thakkinstian {\em et al.} \cite{TH04} compare the combined group of Bb with bb which has the additional benefit of alleviating some small samples sizes that may undermine the assumed normality of the estimated SMDs that is required for the meta-analysis (for more on meta-analysis of SMDs, including normality of the estimates, see, for e.g., Borenstein {\em et al.} \cite{BO&HE&HI&RO09}).  However, some small sizes are still present for the BB group.  The summary statistics for this combined group are also included in Table \ref{table:TH04} and is our focus here.

For comparison of the variances between the BB and Bb/bb groups, we take the ratio of the estimated variances giving, to two decimal places,
$$0.79, 0.01, 0.64, 1.46, 1.82, 1.33, 0.34, 1.33, 2.18, 1.81, 1.01, 1.17, 1.13$$ and we refer to these values as $f_1,\ldots,f_{13}$ respectively.  In the scientific literature it is common to assume that BMD is normally distributed (see, for e.g., \cite{KA02}).  Under this assumption of normality of the underlying populations, if the true variances are equal between the two groups then the distribution of the $i$th ratio is $F_{n_{i1}-1,n_{i2}-1}$ $(i=1,\ldots,13)$ where $n_{i1}$ and $n_{i2}$ are the sample sizes for each of the BB and Bb/bb groups data in the $i$th study.  We reject the possibility of equal variances in a test of variance equality from the $i$th study if $f_i$ is either too small or too large.  Below we provide the $p$-values for each of the 13 tests:
$$0.839, 0.140, 0.358, 0.380, 0.009, 0.084, 0.132, 0.051, 0.000, 0.008, 0.909, 0.622, 0.687$$ with the fifth, sixth, eighth, ninth and tenth studies providing some evidence of unequal variances.  Clearly the smallest of the $p$-values is very small, but this may suggest that a trait specific to this study has resulted in unequal variances.  A question one should ask in such a situation is as to whether a meta-analysis of the SMDs sufficiently robust in regards to departures from equal variances.  In this particular example, it appears that the data across studies is measured on the same scale.  Consequently, the raw mean difference could be used instead of the SMD which is often used not only for interpretative reasons, but also to combine results from studies for which data has been measured on different scales.  When this is not the case, another option is to consider the ratio of means, or to be precise, the log ratio of means \citep{HE99} (when the means are all of the same sign as they are here) or to use Glass's effect which uses the variance estimate from one group only.

\begin{table}[ht]
\centering
\begin{tabular}{llllllll}
  \hline
  &&&&$\rho$&&&\\
Est.   &$.15^2/.15^2$& $.15^2/.2^2$& $.1^2/.2^2$& $.1^2/.3^2$& $.2^2/.15^2$& $.2^2/.1^2$& $.3^2/.1^2$\\ \hline
SMD &0.956 &0.981 &0.998 &1.000 &0.955 &0.924 &0.916\\
MD  &0.961 &0.954 &0.955 &0.957 &0.968 &0.959 &0.942\\
MR  &0.962 &0.953 &0.957 &0.963 &0.966 &0.957 &0.935\\
  \hline
\end{tabular}
\caption{Simulated coverage probabilities for interval estimators of the SMD, Mean Difference (MD) and Mean Ratio (MR) for varying choices of $\rho$.}\label{table:cp}
\end{table}

In Table \ref{table:cp} we provide simulated coverage probabilities (1000 trials) for interval estimators for Cohen's $d$ (SMD), the raw mean difference (MD) and the log ratio of means (MR).  Sample sizes were chosen equal to those for the comparison between the BB and Bb/bb groups where data was sampled from normal distributions with equal means (both 1.0) and $\rho=\sigma_1^2/\sigma_2^2$ shown in the table.  The results were obtained using the \textit{metafor} R package \cite{VI10} with a restricted maximum likelihood estimate for the variance of an assumed random effect.  As can be seen in the table, while the coverage probabilities for the mean difference and log ratio of means are typically close to nominal (0.95), the coverage for the SMD is sensitive to $\rho$ with close to nominal coverage only for two choices of $\rho$, one of which is for equal variances.  This single example highlights some dangers with assuming equal variances and a more exhaustive search for examples will no doubt yield even more evidence.  It should be pointed out that while the use of log ratio of means can be used when study variables are on a different scale and when the means are all of the same sign, in practice it is very common to use Cohen's $d$ and is perhaps even expected depending on the discipline.

\begin{table}[ht]
\centering
\begin{tabular}{llllllll} \hline
$n_1$  & 10& 50& 75 &100 &125 &150 & 190\\
Mean $d$  &        0.505  &0.544 & 0.576 &  0.606 &  0.649 &  0.694 &  0.849 \\
SD $d$ & 0.210 & 0.131 & 0.134 &  0.145 &  0.169  & 0.212  & 0.514 \\ \hline
\end{tabular}\caption{Mean and standard deviations (from 10000 iterations) of simulated Cohen's $d$ for $n_1$ observations generated from $N(1.1,0.12^2)$ $n_2=200-n_1$ from the $N(1.0,0.2^2)$ distribution.}\label{table:d}
\end{table}

The above simulation reported in Table \ref{table:cp} considered only the case when both means are equal.  When they are not equal, interpretation of Cohen's $d$ become difficult since the estimate is highly dependent on the sample sizes when the variances are not equal.  As an example we consider a total sample size of $200 = n_1 + n_2$ and focus on the estimated Cohen's $d$ when $n_1 = 10, 50, 75, 100, 125, 150$ and 190.  For each choice of $n_1$, we generate $n_1$ values from $N(1.1, 0.12^2)$ and $n_2 = 100 - n_1$ values from the $N(1.0, 0.2^2)$ distribution.  This is repeated 10,000 times and in Table \ref{table:d} we report the average and standard deviations of the estimates.  We can see from the reported means that, on average, the estimated $d$s vary from moderate to large depending on the sample sizes.  Additionally, the reported standard deviations also suggest that it would not be unusual to observe small to very large estimates again depending on the sample size allocations.  As this example highlights, interpretation of Cohen's $d$ when the group standard deviations are not equal can be problematic.

Returning to our example data, what we would like to know is whether these results provide persistent evidence of unequal variances in general and as a consequence mean that one should use SMDs with caution, if at all.  Rather than leave this as a subjective problem, we provide a more formal approach to answering this which may benefit researchers by providing at least some formal justification for a lack of evidence or as a means to move away from the assumption of equal variances and adapt the meta-analysis accordingly.  Of course one option is to not use Cohen's $d$ altogether.  However, given its widespread use and interpretation this is unlikely to occur.

\section{Variance stabilization of $F$ statistics}\label{section:var_stab}

An obvious way to query whether the assumption of equal variances is justified across $K$ studies is to consider the ratio of the estimated variances.  However, with sample sizes often varying greatly among studies, interpreting the estimated ratios (both collectively and individually) is not straightforward. Our first focus will therefore be on re-scaling the estimated ratios of variances so that they are all on an easily interpreted scale.

Variance stabilization transformations seek to both normalize and scale an estimate; in this case we are seeking an approximate variance of one.  The benefits can include improved coverage of confidence intervals and a transformed test statistic that is easy to interpret.   Variance stabilization for commonly encountered estimates are still leading to improved inference, as is evidenced by recent research.  For example, Kulinskaya {\em et al.} \cite{KMS10} and Prendergast and Staudte \cite{prst-2014} consider variance stabilization of the difference in binomial proportions using a transformation from Kulinskaya {\em et al.} \cite{KMS-2008}.  When combining evidence from several studies, variance stabilization can be particularly useful given the niceties of dealing with normally distributed statistics.  For some recent examples see \cite{KMS10, MA&PR&ST11,MA&PR&ST13}.  For more discussion on some of the advantages of variance stabilization see Morgenthaler and Staudte \cite{MO&ST12}.  In this section we describe variance stabilization of ratios of estimated variances which is assumed to be $F$-distributed when the data has been sampled from two independent normal distributions.  Simulations will follow in Section \ref{sect:sim}.

Given $S\sim F_{\nu_1,\nu_2}$, the goal is to find a variance stabilizing transformation (VST) $T=T(S)$ for which
$T\sim N(0,1)$ approximately.  For a single study such a transformation is not necessary since probabilities and percentiles from the $F$ distribution are easily obtained.  However, one small advantage in the single study setting is that the transformed statistic is immediately interpretable due to the distribution being free of the degrees of freedom.  This small advantage for a single study leads to a much bigger advantage for multiple studies.   In what follows we explore four possible VSTs that may be used to achieve such advantages.

\subsection{Some possible transformations}\label{section:transformations}

Let $\Phi^{-1}$ denote the inverse cumulative distribution function (inverse cdf) for the standard normal distribution and let $F$ denote the cdf for the $F_{\nu_1,\nu_2}$ distribution.  Then, as pointed out by a referee, a possible transformation is $\Phi^{-1}\left[F(S)\right]\sim N(0,1)$ when $S\sim F_{\nu_1,\nu_2}$.  While this transformation can be obtained computationally, other simple transformations also exist and we discuss these now.

We now consider four competing VSTs of $S\sim F_{\nu_1,\nu_2}$.  For our first two transformations, the technical details are provided in the Appendix.  These transformations are
\begin{eqnarray}
  T_1(S) &=&   \frac {1}{\sqrt {c_1}\,}\;\ln \left(\frac {S}{\e [S]}\right )\, +\frac {\sqrt {c_1}\,}{2}\label{T1}\\
  T_2(S) &=&   \frac {2}{\sqrt {c_2}\,}\;\ln \left(\frac{\sqrt S\,+\sqrt {S+\nu _2/\nu _1}\,}
  {\sqrt {\e[S]}\,+\sqrt {\e[S]+\nu _2/\nu _1}\,} \right)
    + \frac{\sqrt {c_2}\,(2\nu _1+\nu _2-2)}{4\sqrt {\nu _1^2+2\nu _1\nu _2-2\nu _1}\,}  ~\label{T2}
\end{eqnarray}
where $c_1=2(\nu_1+\nu _2-2)/[\nu _1(\nu _2-4)]$, $c_2=2/(\nu _2-4)$ and $\e[S]=\nu_2/(\nu_2-2)$ is the mean for the $F_{\nu_1,\nu_2}$ distribution for $\nu_2>2$.  Both of these transformations have approximate mean zero and variance one.

The next transformation that we consider is Paulson's normalizing transformation \cite{PA42} of the $F_{\nu _1,\nu _2}$ statistic given as
\begin{equation}\label{T3}
    T_3(S) = \left \{\left (1-\frac {2}{9\nu_2}\right)S^{1/3}-\left (1-\frac {2}{9\nu_1}\right )\right \}
           \left \{\left (\frac {2S^{2/3}}{9\nu_2}+\frac {2}{9\nu_1}\right )\right \}^{-1/2}~.
\end{equation}

As was the case with $T_1$ and $T_2$,  a re-centering of $T_3$ could also be carried out so that its expected value is closer to zero.  However, for this transformation the adjustment is very small and our simulations reveal it is not necessary.  In fact this transformation as it is is quite remarkable in that when compared to $\Phi^{-1}\left[F(S)\right]$, the results are almost identical so that $T_3$ provides a simple alternative.  For example, when simulating 1000 observations from the $F_{\nu_1,\nu_2}$ distribution for all choices of $\nu_1$ and $\nu_2$ in $5,\ldots,100$, the minimum correlation between the transformed data using each of the transformations was 0.99969.

The final transformation we now describe a slight variation of the VST introduced in Chapter 23 of Kulinskaya {\em et al.} \cite{KMS-2008}.  While it is more complicated than the other transformations, an advantage it has over $T_1$ and $T_2$ is that it exists for all degrees of freedom.  It is also more applicable in general than the other transformations since it can be applied to non-central $F$ distributed test statistics.  Consider a test statistic $S_\lambda$ for which $S_\lambda\sim F_{\nu_1,\nu_2}(\lambda )$ where $\lambda$ is the non-centrality parameter.  Let $\Delta = F^{-1}_{\nu_1,\nu_2}(0.5)$ where $F^{-1}_{\nu_1,\nu_2}$ denotes the inverse cumulative distribution function so that $\Delta$ is the median of the $F_{\nu_1,\nu_2}$ (i.e. with $\lambda=0$; the central $F$) distribution.  Let
\begin{equation}
S^*_\lambda = \left\{\begin{array}{ll} S_\lambda, & S_\lambda > \Delta\\
  F^{-1}_{\nu_1,\nu_2}[1- F_{\nu_1,\nu_2}(S_\lambda)] & S_\lambda \leq \Delta\end{array}\right.
\end{equation}
where $F_{\nu_1,\nu_2}$ is the cumulative distribution function of $S_\lambda$.

For $\text{sign}(S_\lambda ;\Delta)=1$ if $S_\lambda > \Delta$ and $-1$ otherwise, the transformation function is
\begin{align}
T_4(S_\lambda ;\nu_1,\nu_2)=&\left(\frac{\nu_2}{\nu_2 + 1}\right)\text{sign}(S_\lambda;\Delta)\times \sqrt{\frac{\nu_2}{2}}\Bigg[\text{cosh}^{-1}\left(\frac{\nu_1S_\lambda^*+\nu_2}{\sqrt{\nu_1\nu_2\Delta+\nu_2^2}}\right)\nonumber\\
&\;\;\;\;\;\;\;\;\;\;\;\;\;\;-\text{cosh}^{-1}\left(\sqrt{\nu_1\Delta/\nu_2+1}\right)\Bigg]~.\label{T4}
\end{align}
 This VST is that of Kulinskaya {\em et al.} \citep[p.199]{KMS-2008}, but multiplied by  $\nu_2/(\nu_2 + 1)$.  We have used this adjustment because our extensive simulations reveal that it improves performance when $\nu_2$ is small. Kulinskaya {\em et al.} \cite{KMS-2008} show that $h(S;\nu_1,\nu_2)$ is approximately normally distributed with variance one  even when the degrees of freedom are small, and that it has mean 0 under the null and mean increasing in $\lambda $ under alternatives.  They also provide applications for the ANOVA $F$-tests.

 While we have referred to these transformations as VSTs, this is strictly only true in general for $T_1$ for any $\rho$.  The other transformations are VSTs for the special case of $\rho=1$ although the transformations are still useful for any $\rho$ in what follows.

\subsection{Q-Q plots to assess violations of the assumption of equal variances}

Returning to our meta analysis of SMD's where we want a meta-analysis of variance ratios, assume that we have $K$ independent $F$-distributed random variables denoted $S_1,\ldots,S_K$ with potentially different degrees of freedom (denoted $\nu_{k1}$ and $\nu_{k2}$ for the $k$th study) and population variances $\sigma_{k1}^2$ and $\sigma_{k2}^2$.  When $\sigma_{k1}^2\neq\sigma_{k2}^2$, then the $S_k$s are distributed as $(\sigma_{k1}^2/\sigma_{k2}^2)F_{\nu_1,\nu_2}$ random variables.

Below let $T_i$ denote one of the transformations $T_1$, $T_2$, $T_3$ and $T_4$ given in Section \ref{section:transformations}.   We have
\begin{equation}
Z_{k} = T_i(S_k;\nu_{k1},\nu_{k2}) \asim N(0, 1) \;\;\;\text{if $\sigma_{k1}^2=\sigma_{k2}^2$ for each $k$}\nonumber
\end{equation} where the closeness of the approximation may depend on the VST and the degrees of freedom.  Since the $S_k$'s are independent, if the variances within the $K$ studies are in fact equal, then the ordered $Z_k$'s should resemble data randomly generated from a standard normal distribution.  Let $Z_{[1]}\leq \ldots \leq Z_{[K]}$ denote the ordered $Z_k$'s.  Then one could visually assess the possibility of violations of the equal variance assumption by utilizing Quantile-Quantile (Q-Q) plots of the
$$Z_{[k]}\;\;\text{versus}\;\;\Phi^{-1}\left(\frac{k-0.5}{K}\right)$$ where $\Phi^{-1}(p)=z_p$ is the $p\times 100$th percentile of the standard normal distribution such that $P(Z\leq z_p)=p$ for $Z\sim N(0,1)$.

\begin{figure}[h!t]
\centering
\includegraphics[scale=0.8]{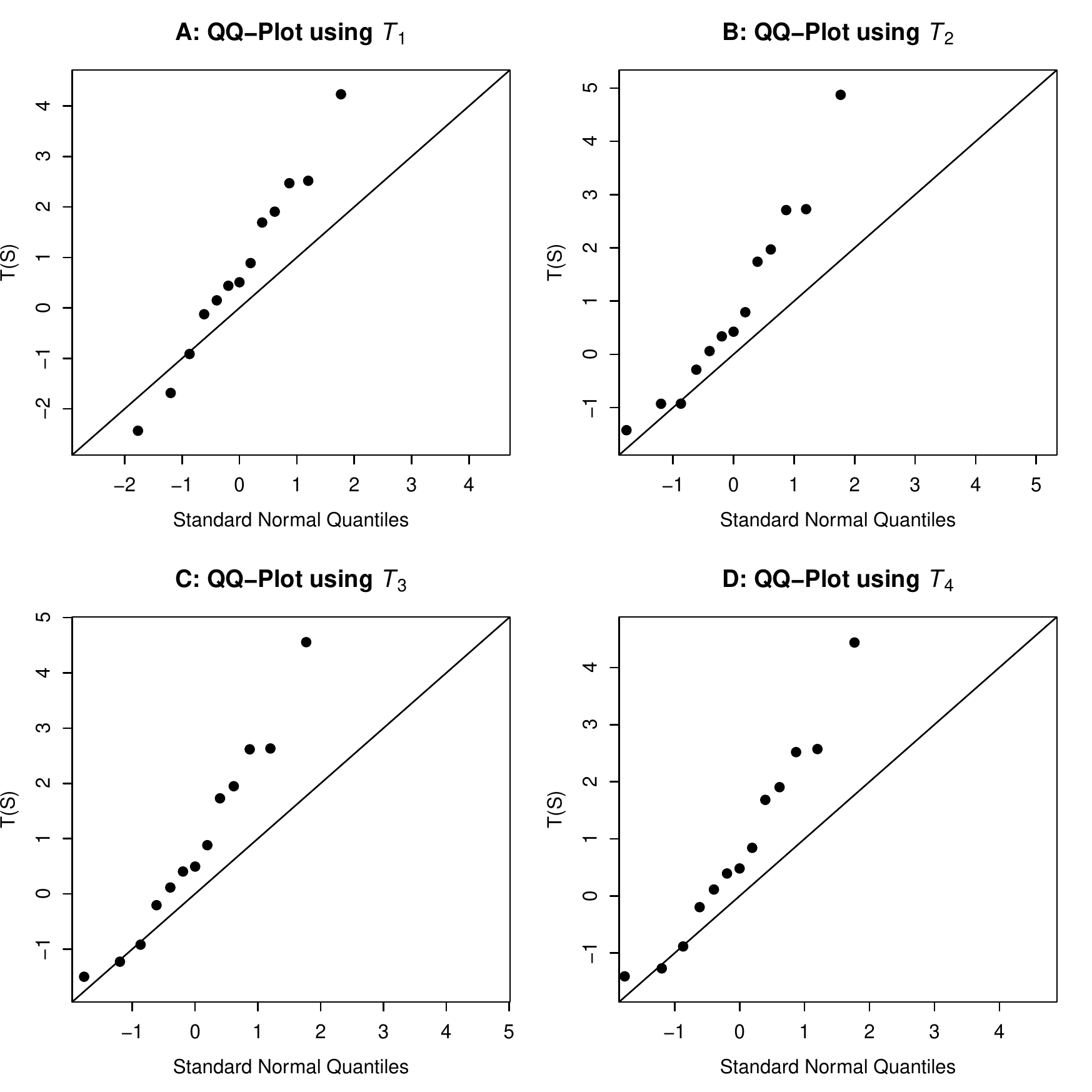}
\caption{Q-Q plots of the ordered VST transformed ratios for the BB vs Bb/bb data in Table \ref{table:TH04}.  Plots A, B, C and D are for the transformations in \eqref{T1}, \eqref{T2}, \eqref{T3} and \eqref{T4} respectively.}\label{figure1}
\end{figure}

Returning to our motivating example of Section \ref{section:mot}, in Figure \ref{figure1} we provide the Q-Q plots for each of the transformations $T_1,\ldots,T_4$.  All of the plots are similar with only some minor differences and a more in depth comparison of the VSTs and their ability to achieve approximate standard normality will be explored in greater depth later.  All of the Q-Q plots suggest potential problems with the assumption of equal variances with a trend towards  larger than expected $Z_k$'s.  While these Q-Q plots are informative and easy to obtain, in this example and others some researchers would like an accompanying test for unequal variances."  This is introduced in the next section.  For those who prefer to interpret a meta-analysis of  effect sizes, in this case
the ratio of variances, one can skip to the following section.  Both methodologies are based on the same  combination
of transformed F-test statistics

\subsection{Omnibus tests for unequal variances}\label{sect:test_for_unequal}

Continuing with the notation introduced in the previous section, two meta-analytic test statistics that may be used to combine the evidence are
\begin{equation}
Z=\sqrt{K}\times \overline{Z}=\frac{1}{\sqrt{K}}\sum^K_{k=1}Z_{k} \asim N(0, 1)\label{meanZt}
\end{equation}
and
\begin{equation}
X^2 = \sum^K_{k=1}Z_k^2 \asim \chi^2_{K}\label{sumZsq}
\end{equation}
to test for $H_0:\sigma_{k1}^2=\sigma_{k2}^2$ for every $k=1,\ldots, K$.

The two test statistics above apply an equal weighting for each study.  However, another possibility would be to apply a greater weighting to studies with smaller estimator variability on the non-transformed scale.  Hence, weighted versions of the above are
\begin{equation}
Z_w=\frac{1}{\sqrt{\sum^K_{k=1}w_k^2}}\times \sum^K_{k=1}w_kZ_{k} \asim N(0, 1)\label{meanZw}
\end{equation}
and
\begin{equation}
X_w^2 = \sum^K_{k=1}w_kZ_k^2\label{sumZsqw}
\end{equation}
where $w_1,\ldots,w_K$ are nonzero weights where $\sum^K_{k=1}w_k=1$.  Under the assumption of $\sigma_{k1}^2=\sigma_{k2}^2$ the distribution of $X_w^2$ is not straightforward although some approximating distributions do exist \cite[see, for e.g.,][]{YU&BE10}.  However, for what we require shortly, it is quite efficient to simply use a Monte-Carlo simulation to accurately obtain what is needed.  In terms of the weights there are many possibilities and we leave further guidance on this until the next section.

As expected, if the population variances are not equal across the studies, then the evidence against equal variances is expected to grow with increasing $K$ (the number of studies).  Note that there is also no requirement that $\sigma^2_{k1}\neq \sigma^2_{k2}$ across all $k=1,\ldots,K$ simultaneously.
Given that the distribution of $Z$ in \eqref{meanZt} (and also $Z_w$)  does not depend on the number of studies, it is immediately interpretable as a measure of evidence against equal variances and, if $Z>0$, in favor of
 $\sigma_{k1}^2> \sigma_{k2}^2$ for all $k$ with inequality for at least one $k$ (or similarly for $\sigma_{k1}^2< \sigma_{k2}^2$ if $Z<0$). However, one is usually
 not willing or able to restrict the alternative to such a  \lq one-sided\rq\ possibility, so the second test
 $\chi ^2$ statistic may be more appropriate.
 In a meta-analysis  it is widely believed that differences in studies might lead to different study-parameters of interest and random effects can be used to describe such differences.  It is therefore possible that differences in studies could result in different magnitudes of ratios of variances.  Therefore the test statistic in \eqref{sumZsq} is more powerful when some studies have ratios greater than one and others less than one.  A $p$-value for testing against the hypothesis of equal variances can be computed using $P(\cal{X} > X^2)$ where $\cal{X}\sim\chi^2_{K}$.  A $p$-value associated with the test statistic in \eqref{sumZsqw} is not exactly determinable.  However, to obtain an accurate approximation via simulation, one simply needs to randomly generate $M$ $X_w^2s$ under the null using $\mathbf{X}\mathbf{w}$ where $\mathbf{X}$ is an $M\times K$ matrix of randomly generated $\chi^2_1$ values and $\mathbf{w}$ is the column vector of weights.  The simulated $p$-value is then simply the proportion of values in the resulting vector that are greater than the observed test statistic.

\subsection{Estimates of an assumed fixed ratio of variances}\label{section:estimates}

In this section we provide our first insights into estimating an assumed fixed ratio of variances.  While the following estimators are motivated by the VSTs, it should be noted that there is strong link with Maximum Likelihood Estimators (MLEs) that we consider in the next section.  The MLEs also allow for the simple introduction of a random effect and this will also be considered in the next section.

Suppose that $\sigma^2_{k1}/\sigma_{k2}^2 = \rho$ for all $k = 1,\ldots,K$.  That is, the ratio of variances is assumed fixed across all studies where $\rho =1$ is assumed when utilizing SMDs.  In this section we consider combining evidence across all studies to obtain point and interval estimates for $\rho$.  Under the assumption that data are sampled from independent normal distributions, we have that $S_k/\rho \sim F_{n_{k1}-1,n_{k2}-1}$ so that, from Section \ref{sect:test_for_unequal},
\begin{equation}
T_i(S_k/\rho;\nu_{k1},\nu_{k2}) \asim N(0, 1)\nonumber
\end{equation}
and subsequently
\begin{equation}
Z_{iw}=\frac{1}{\sqrt{\sum^K_{k=1}w_k^2}}\times \sum^K_{k=1}w_kT_i(S_k/\rho;\nu_{k1},\nu_{k2}) \asim N(0, 1).\label{meanZw2}
\end{equation}

Let $c_{k1}$ denote the $k$th study specific $c_1$ used in \eqref{T1} and also let $\e_0(S_k)=\e(S_k/\rho)=\nu_{k2}/(\nu_{k2} - 1)$ denote the expected value of $S_k$ under the null hypothesis of equal variances and which is free of $\rho$.  Consequently, $Z_{iw}$ is a pivotal quantity for $\rho$ and an approximate $(1-\alpha)\times 100$\% confidence interval for $\rho$ can be obtained by solving
\begin{equation*}
|Z_{iw}|= z_{1-\alpha/2}
\end{equation*}
where $z_{1-\alpha/2}$ is the $(1-\alpha/2)100$\% percentile of the $N(0,1)$ distribution.

For illustrative purposes we shall focus our attention on $T_1$ from \eqref{T1}.  For this simple VST, formulae for the interval estimates of $\rho$ can be derived as closed-form expressions which will aid in discussion.  Estimates for the other transformations can be obtained through simple computational root-solving.  Using the definition of $T_1$, simple rearranging of \eqref{meanZw2} leads to
\begin{equation}
\left(\sum^K_{k=1}\frac{w_k}{\sqrt{c_{k1}}}\right)^{-1}\sum^K_{k=1}\frac{w_k}{\sqrt{c_{k1}}}\left[\ln(S_k) + \ln\left(\frac{n_{k2}-3}{n_{k2} - 1}\right) + \frac{c_{k1}}{2}\right]\asim N[\ln(\rho), V]\label{ln_rho_hat}
\end{equation}
where $$V = \left(\sum^K_{k=1}\frac{w_k}{\sqrt{c_{k1}}}\right)^{-2}\sum^K_{k=1}w_k^2.$$  Therefore, the left hand side of \eqref{ln_rho_hat} is an estimator of $\ln(\rho)$ and which has variance $V$.  Consequently, we denote this estimator as $\widehat{\ln(\rho)}$ and a $(1-\alpha)\times 100$\% confidence interval for $\ln(\rho)$ is $\widehat{\ln(\rho)}\pm z_{1-\alpha/2} \sqrt{V}$.  Finally, we exponentiate to obtain our confidence interval for $\rho$ given as
\begin{equation}
\big(\exp[\widehat{\ln(\rho)}- z_{1-\alpha/2} \sqrt{V}],\ \exp[\widehat{\ln(\rho)}+ z_{1-\alpha/2} \sqrt{V}]\big) \label{ci_rho}
\end{equation}
where $\ln(\rho)$ and $V$ are given in \eqref{ln_rho_hat}.  This confidence interval is simple to compute which is an advantage for the $T_1$ VST.  For the other VSTs, the estimates can be obtained computationally which, using a package such as R, is a simple task.

So far in this section we have not provided any guidance as to suitable choices for the weights $w_1,\ldots,w_K$.  In the next section, we show that choice of \begin{equation}
w_k=\frac{1}{\sqrt{c_{k1}}},\;\;k=1,\ldots,K\label{weights}
 \end{equation}
 in \eqref{ln_rho_hat} results in an MLE estimator of $\rho$ (with an equivalent variance $V$) based on the log of the ratio of sample variances.  This will therefore provide motivation for the choice of weights.  However, future work may include different choices of weights that may, for example, be used to provide some degree of robustness in the estimation to protect against outliers.

\section{Meta-analysis based on maximum likelihood estimation}\label{section:MLE}

In the previous section we considered meta-analysis for the ratio of variances following a variance stabilization of the $F$-test statistics.  Another possibility is to consider MLEs for the ratio of variances.  We will firstly consider MLE estimation based on the true distribution of the ratio of sample variances which allows for estimation of an assumed fixed ratio across studies.  We will then consider an extension to allow for a simple random effect on the ratio with estimation based on the approximate normal distribution for the log of the ratio of sample variances.  Throughout we will continue to use the notation of the previous sections.

\subsection{MLE estimation based on the re-scaled $F$ distribution}\label{sect:MLE_F}

Let $g(f;\nu_1,\nu_2)$ denote the probability density function for the $F$ distribution with degrees of freedom $\nu_1$ and $\nu_2$.  Then, under the assumption that the data are sampled from independent normal distributions, it is simple to verify that the probability density function for $S_k$ is
\begin{equation}
f(s) = \frac{1}{\rho_k}g(s/\rho_k;n_{k1}-1,n_{k2-1})
\end{equation}
where $\rho_k=\sigma_{k1}^2/\sigma_{k2}^2$.

It is not difficult to obtain the point and interval MLE estimates for a fixed $\rho_k=\rho$ $(k=1,\ldots,K)$ based on the likelihood function above when using routine optimization functions within a package such as R that includes a computationally computed Hessian matrix.

\subsection{MLE estimation based on the approximate normal distribution}

The simplistic form of $T_1$ from Section \ref{section:var_stab} provides an opportunity to obtain estimates for the ratio of variances based on approximate normality of the log of the ratio of sample variances.  For simplicity throughout let $\omega = \ln(\rho)$

\subsubsection{Fixed effect model}\label{section:MLE_norm}

Using the fact that $T_1(S_k)$ is approximately $N(0,1)$ distributed we have
\begin{equation}
\ln(S_k) \asim N(\mu_k, c_{1k})\label{ln_k_norm}
\end{equation}
where $\mu_k=\omega+\ln\left\{\nu_{2k}/(\nu_{2k}-2)\right\}-c_{k1}/2$ and where $c_{k1}$ is defined as in $c_1$ in \eqref{T1}, but specific to the $k$th study.

Under the assumption of a common fixed $\rho=\rho_k$ across all studies,  the log-likelihood function obtained from \eqref{ln_k_norm} and ignoring constant terms is
\begin{equation}
l(s_1,\ldots,s_K;\omega) = -\frac{1}{2}\sum^K_{k=1}\frac{1}{c_{k1}}\left[\ln(s_k)-\omega-\ln\left\{\nu_{2k}/(\nu_{2k}-2)\right\}+c_{k1}/2\right]^2.\label{ll}
\end{equation}

Solving for $l'(s_1,\ldots,s_K;\ln(\rho))=0$ for $\ln(\rho)$ gives the MLE estimate
\begin{equation}
\widehat{\omega}_{MLE} = \left(\sum^K_{k=1}\frac{1}{c_{k1}}\right)^{-1}\sum^K_{k=1}\frac{1}{c_{k1}}\left[\ln(s_k) + \ln\left(\frac{n_{k2}-3}{n_{k2} - 1}\right) + \frac{c_{k1}}{2}\right]
\end{equation}
which is identical to the estimate obtained from \eqref{ln_rho_hat} when the weights, $w_{k}s$, are chosen to be $1/\sqrt{c_{k1}}$.  A closed form solution for the variance of the MLE also exists and is given as
\begin{equation*}
V_{MLE}^*=\frac{1}{-l''(s_1,\ldots,s_K;\omega)}=\left(\sum^K_{k=1}\frac{1}{c_{k1}}\right)^{-1}
\end{equation*}
which is equivalent to $V$ from \eqref{ln_rho_hat} when the weights are chosen to be $1/\sqrt{c_{k1}}$.  This leads to an approximate $(1-\alpha)\times 100$\% confidence interval for $\rho$ as
\begin{equation}
\left(\exp[\widehat{\omega}_{MLE}- z_{1-\alpha/2} \sqrt{V_{MLE}^*}],\ \exp[\widehat{\omega}_{MLE}+ z_{1-\alpha/2} \sqrt{V_{MLE}^*}]\right). \label{ci_rho_MLE}
\end{equation}

\subsubsection{Random effects model}\label{sect:REM}

As is done for classic random effects models for effect sizes, instead of assuming a fixed $\rho$ across all studies, we can instead introduce a random effect to explain unmeasurable heterogeneity among the $\rho_k$s.  From \eqref{ln_k_norm}, a simple random effect model assumes
\begin{equation}
\ln(S_k) \asim N(\mu_k, c_{1k}+\tau^2)\label{ln_k_norm_rand}
\end{equation}
where $\tau^2$ is the variance of an assumed central normal additive random effect for $\ln(S_k)$.  While a closed-form solution for the MLEs of $\rho$ and $\tau$ have not been obtained, computationally these are not too difficult to achieve due to the niceties of dealing with the normal distribution.

For $Y_k=\ln(S_k)\sim N(\mu_k,c_{1k}+\tau^2)$, ignoring constants, a log-likelihood function is
\begin{equation}
l(y_1,\ldots,y_K;\omega,\tau)=-\frac{1}{2}\sum^{K}_{k=1}\left[\ln(c_{1k}+\tau^2)+\frac{(y_k-\mu_k)^2}{c_{1k}+\tau^2}\right].\label{l}
\end{equation}

Taking the first derivative of \eqref{l} with respect to both $\omega$ and $\tau^2$, the MLEs for each (denoted $\widehat{\omega}_{MLE}$ and $\widehat{\tau^2}_{MLE}$)  are the solutions to
\begin{equation*}
\frac{\partial l}{\partial \omega}=\sum^K_{k=1}\frac{y_k-\mu_k}{c_{1k}+\tau^2}=0\;\;\text{and}\;\;
\frac{\partial l}{\partial \tau^2}=\frac{1}{2}\sum^K_{k=1}\left[\frac{(y_k-\mu_k)^2}{(c_{1k}+\tau^2)^2}-\frac{1}{c_{1k}+\tau^2}\right]=0.
\end{equation*}
To assure that the estimate to $\tau$ is positive, we found that the best approach was to re-write the likelihood and estimate the parameter $\ln(\tau)$.  Exponentiating was then used to achieve the estimate for $\tau$.

We can also derive approximate variances for the MLEs (details are in Appendix \ref{appendix:SE}),
\begin{equation}
V_{MLE,\omega}=\left(\sum^K_{k=1}\frac{1}{c_{k1} + \widehat{\tau^2}_{MLE}}\right)^{-1},\;\;\;
V_{MLE,\tau^2}=2\left(\sum^K_{k=1}\frac{1}{(c_{k1} + \widehat{\tau^2}_{MLE})^2}\right)^{-1}.
\end{equation}

When constructing intervals using the appropriate percentile from the normal distribution, our simulation studies (to be discussed in the next section) showed that the intervals were too narrow.  Therefore our suggested approximate $(1-\alpha)\times 100$\% confidence interval for $\rho$ is
\begin{equation}
\left(\exp[\widehat{\omega}_{MLE}- t_{K-1,1-\alpha/2} \sqrt{V_{MLE,\omega}}],\ \exp[\widehat{\omega}_{MLE}+ t_{K-1,1-\alpha/2} \sqrt{V_{MLE,\omega}}]\right) \label{ci_rho_MLE2}
\end{equation}
where $t_{K-1,1-\alpha/2}$ is the $(1-\alpha/2)\times 100$\% percentile from the $t_{K-1}$ distribution which results in intervals with improved coverage.

\section{Simulation Studies}\label{sect:sim}

In this section we consider several simulation studies, the first of which are focused on the performance of the transformation to achieve approximate standard normality followed by assessment of performance when applied in a meta-analysis setting.

\subsection{Simulation 1:\quad  Approximate standard normality of the transformed $F$-statistics under the assumption of equal variances}\label{sect:sim_size}

In Section \ref{section:var_stab} we noted that a correction to the $T_4$ transformation function can be useful to improve achieving approximate standard normality.  This was discovered by considering contour plots of the size of the tests (tail probabilities) from Section \ref{sect:test_for_unequal} under the assumption of equal variances.  We provide some examples of these plots here for $T_1$, $T_2$, $T_3$ and $T_4$.

\begin{figure}[h!t]
\centering
\includegraphics[scale=0.8]{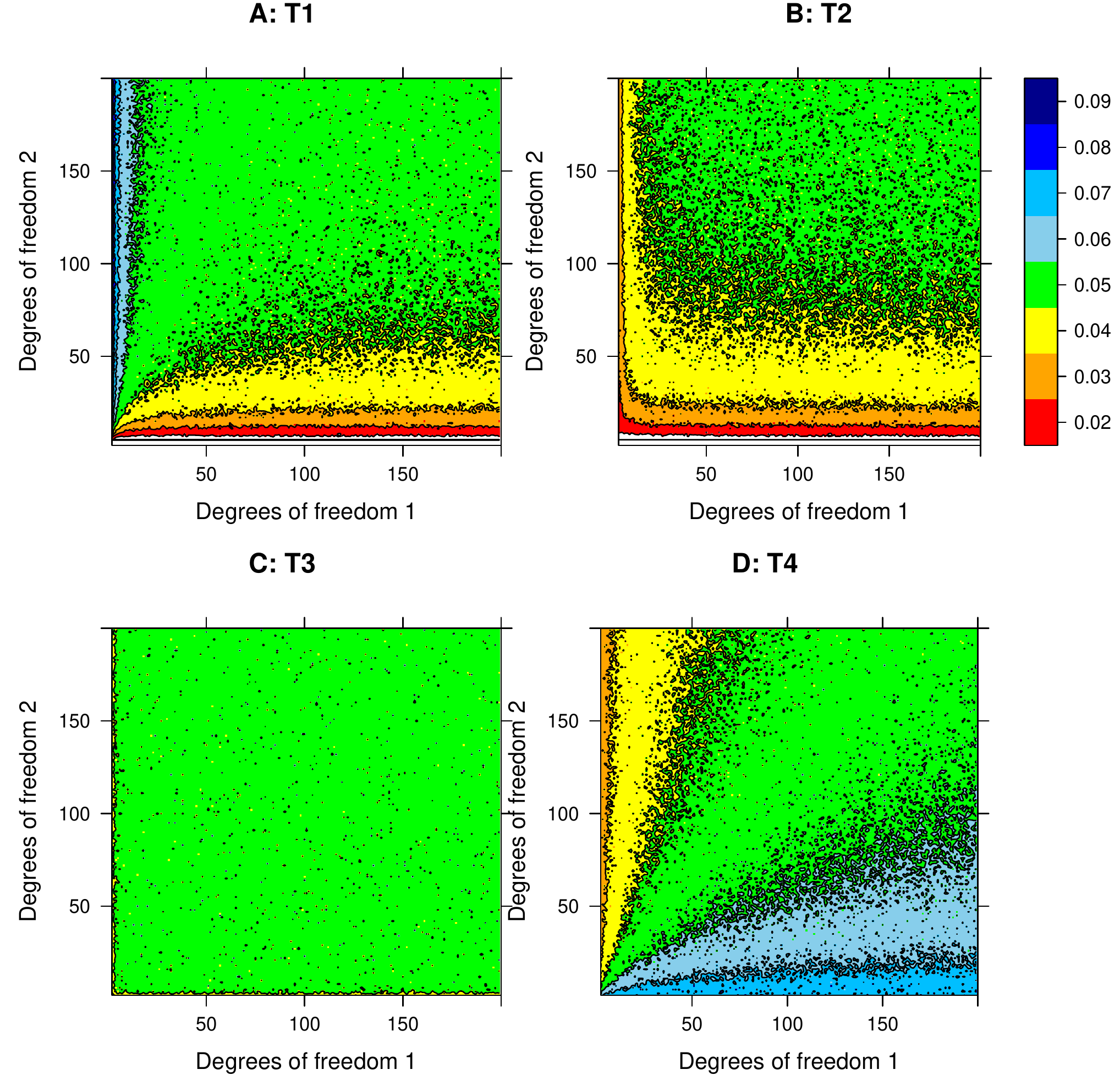}
\caption{Contour plots for the empirical size of the test following variance stabilization when the nominal size is $\alpha = 0.05$.  Plots A, B, C and D are for the transformations in \eqref{T1}, \eqref{T2}, \eqref{T3} and \eqref{T4} respectively. 10,000 iterations were used in the simulation for each combination of degrees of freedom.}\label{figure2}
\end{figure}

The simulated size of the test is depicted which is the proportion of times that $Z^2$ exceeded the cutoff $1.96^2$ for a nominal size of $\alpha=5$\% where $Z$ is from \eqref{meanZt} but for a single study.  That is, it is simply the VST transformed test statistics using the VSTs $T_1$ (Plot A) through to $T_4$ (Plot D).  While all the transformations perform well across most of the degrees of freedom combinations (since the empirical size is close to nominal), the Paulson transformation $T_3$ is quite remarkable in the sense that it consistently results in an excellent empirical size across all of the choices.  As noted previously, $T_3(S)$ is almost identical to $\Phi^{-1}[F(S)]$ for $S\sim F_{\nu_1,\nu_2}$. The white space in Plots A and B are for when the transformations  are not defined for $T_1$ and $T_2$ which occurs when $\nu_2\leq 4$.  VST $T_4$ is defined across all degrees of freedom and therefore does not suffer from this same problem, but for consistency $T_3$ is the overall best performer when it comes to size.  It should be pointed out that when  the degrees of freedom is small, transformations such as $T_1$ are sensitive to which group is used in the denominator and which is used in the numerator for the ratio.  So from a purely testing point of view, $T_3$ is the safest option.  However, this is for a single study only and in a meta analysis the problem is likely to only be problematic if there is a persistent small sample size in one group across several studies.

Similar results were also found when we considered sizes of $\alpha=0.01$ and $\alpha=0.1$ (not shown).  While all of the VSTs performed well, it was the Paulson transformation that consistently provided excellent results across all combinations for the degrees of freedom.

We now consider the performance of the transformation function  in achieving standard normality over varying $\nu_1$ and $\nu_2$.

\begin{figure}[h!t]
\centering
\includegraphics[scale=1.0]{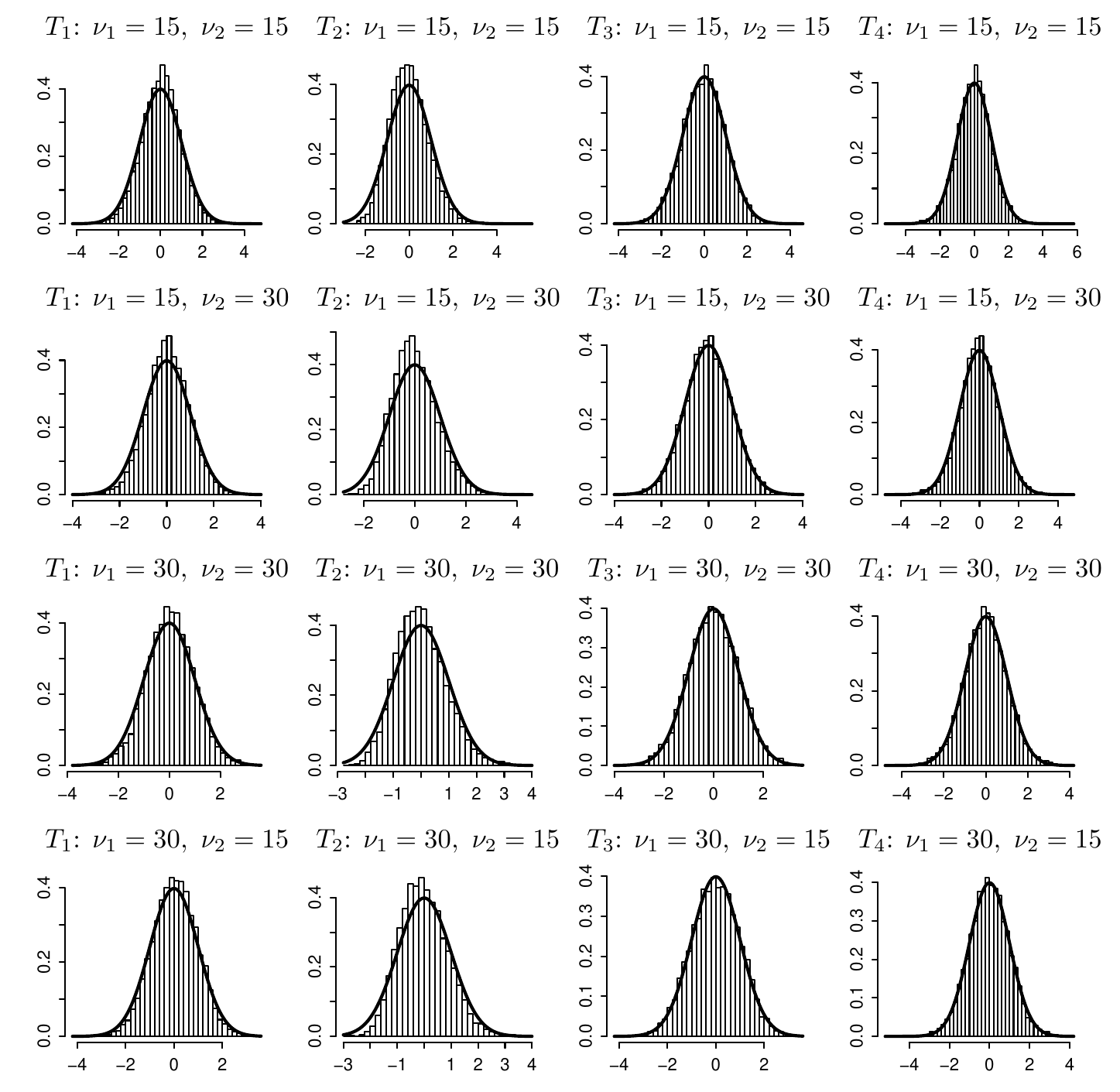}
\caption{Histograms of 10,000 simulated values from the $F_{\nu_1,\nu_2}$ distribution following transformation using each of the VSTs $T_1,\ldots,T_4$.  The black line denotes the standard normal probability density function for comparison.}\label{figure3}
\end{figure}

In Figure \ref{figure3} we provide histograms of 10,000 observations generated form the $F_{\nu_1,\nu_2}$ distribution after application of each of VSTs $T_1$, $T_2$, $T_3$ and $T_4$.  Some combinations of $\nu_1$ and $\nu_2$ are from $\{15,30\}$ and the black line depicts the standard normal density function.  All VSTs achieve at least approximate normality for these combinations of degrees of freedom.  However, it is the $T_3$ and $T_4$ VSTs that are the best performers followed by $T_1$ provided both degrees of freedom are not too small.  We also tried smaller degrees of freedom values (e.g. combinations from 10,20; not shown) and the performance of $T_2$ diminished the most.  $T_1$ still did a reasonable job of normalizing the data and $T_3$ and $T_4$ performed exceptionally well.  When increasing the degrees of freedom all methods did very well in normalizing $F$-statistics.

\subsection{Simulation 2:\quad Meta-analysis for estimation of $\rho$.}\label{sect:meta_rho}

In this section we consider a meta-analysis for the estimation of the ratio variances $(\rho)$.  We consider both an assumed fixed $\rho$ across all studies and also a random effect model that allows $\rho$ to vary across the studies.  Simulations are used to obtain approximate values for bias, actual coverage probability (CP) and confidence interval width (W) for $\rho$ as well as the bias of the variance term associated with the random effect for the log of the ratio of variances.

\setlength{\tabcolsep}{4pt}
\begin{table}[h!t]
\centering
\begin{small}
 \begin{tabular}{cccccccccccccccccc}
   \cmidrule{3-18}
&& \multicolumn{3}{c}{$T_1$}& & \multicolumn{3}{c}{$T_3$}& & \multicolumn{3}{c}{$FE$}& & \multicolumn{4}{c}{$RE$} \\
$\tau$&$\rho$&Bias&CP&W&&Bias&CP&W&&Bias&CP&W&&Bias&CP&W&Bias$_\tau$\\
\cmidrule{1-5} \cmidrule{7-9} \cmidrule{11-13} \cmidrule{15-18}
0.0& 0.20 & 0.00 & 0.95 & 0.06 && 0.00 & 0.96 & 0.06 && 0.00 & 0.96 & 0.06 && 0.00 & 0.98 & 0.07 & 0.05 \\
   & 0.50 & 0.00 & 0.95 & 0.15 && 0.00 & 0.95 & 0.15 &&-0.00 & 0.95 & 0.15 && 0.00 & 0.97 & 0.17 & 0.04 \\
   & 0.80 & 0.01 & 0.95 & 0.24 && 0.01 & 0.96 & 0.24 && 0.01 & 0.96 & 0.24 && 0.01 & 0.98 & 0.28 & 0.05 \\
   & 1.00 & 0.01 & 0.95 & 0.30 && 0.01 & 0.95 & 0.30 && 0.01 & 0.95 & 0.30 && 0.01 & 0.97 & 0.35 & 0.04 \\
   & 1.25 & 0.00 & 0.96 & 0.37 && 0.00 & 0.95 & 0.37 &&-0.00 & 0.95 & 0.37 &&-0.00 & 0.98 & 0.45 & 0.06 \\
   & 1.50 & 0.00 & 0.95 & 0.44 && 0.00 & 0.95 & 0.44 &&-0.00 & 0.95 & 0.44 &&-0.00 & 0.97 & 0.54 & 0.06 \\
   & 2.00 &-0.00 & 0.94 & 0.59 &&-0.00 & 0.95 & 0.59 &&-0.01 & 0.95 & 0.59 &&-0.01 & 0.97 & 0.71 & 0.05 \\
   & 5.00 & 0.02 & 0.95 & 1.48 && 0.02 & 0.95 & 1.48 && 0.01 & 0.96 & 1.47 && 0.01 & 0.98 & 1.77 & 0.05 \\  \cmidrule{1-5} \cmidrule{7-9} \cmidrule{11-13} \cmidrule{15-18}
0.2& 0.20 & 0.00 & 0.82 & 0.06 && 0.00 & 0.83 & 0.06 && 0.00 & 0.83 & 0.06 && 0.00 & 0.94 & 0.08 &-0.05 \\
   & 0.50 & 0.01 & 0.82 & 0.15 && 0.01 & 0.83 & 0.15 && 0.01 & 0.81 & 0.15 && 0.00 & 0.93 & 0.21 &-0.05 \\
   & 0.80 & 0.00 & 0.83 & 0.24 && 0.01 & 0.84 & 0.24 && 0.01 & 0.83 & 0.24 && 0.00 & 0.94 & 0.34 &-0.06 \\
   & 1.00 & 0.00 & 0.83 & 0.29 && 0.01 & 0.83 & 0.30 && 0.01 & 0.83 & 0.30 && 0.00 & 0.94 & 0.43 &-0.05 \\
   & 1.25 &-0.00 & 0.83 & 0.37 && 0.00 & 0.83 & 0.37 && 0.01 & 0.84 & 0.37 &&-0.00 & 0.94 & 0.53 &-0.05 \\
   & 1.50 & 0.02 & 0.82 & 0.44 && 0.02 & 0.82 & 0.45 && 0.03 & 0.82 & 0.45 && 0.01 & 0.93 & 0.64 &-0.05 \\
   & 2.00 & 0.01 & 0.84 & 0.59 && 0.02 & 0.84 & 0.60 && 0.03 & 0.84 & 0.60 && 0.01 & 0.94 & 0.84 &-0.05 \\
   & 5.00 & 0.03 & 0.83 & 1.48 && 0.04 & 0.83 & 1.49 && 0.07 & 0.83 & 1.49 && 0.03 & 0.94 & 2.10 &-0.06 \\  \cmidrule{1-5} \cmidrule{7-9} \cmidrule{11-13} \cmidrule{15-18}
0.4& 0.20 & 0.00 & 0.63 & 0.06 && 0.01 & 0.62 & 0.06 && 0.01 & 0.61 & 0.06 && 0.00 & 0.93 & 0.12 &-0.05 \\
   & 0.50 & 0.01 & 0.61 & 0.15 && 0.02 & 0.61 & 0.15 && 0.03 & 0.59 & 0.16 && 0.01 & 0.94 & 0.31 &-0.04 \\
   & 0.80 & 0.01 & 0.63 & 0.24 && 0.02 & 0.64 & 0.24 && 0.04 & 0.63 & 0.25 && 0.00 & 0.94 & 0.50 &-0.04 \\
   & 1.00 & 0.02 & 0.63 & 0.30 && 0.03 & 0.63 & 0.31 && 0.06 & 0.60 & 0.31 && 0.01 & 0.94 & 0.62 &-0.05 \\
   & 1.25 & 0.00 & 0.65 & 0.37 && 0.02 & 0.65 & 0.38 && 0.06 & 0.62 & 0.39 && 0.00 & 0.94 & 0.78 &-0.04 \\
   & 1.50 & 0.02 & 0.62 & 0.45 && 0.04 & 0.62 & 0.46 && 0.08 & 0.61 & 0.47 && 0.02 & 0.92 & 0.92 &-0.06 \\
   & 2.00 & 0.03 & 0.61 & 0.60 && 0.06 & 0.62 & 0.61 && 0.11 & 0.60 & 0.63 && 0.01 & 0.93 & 1.24 &-0.04 \\
   & 5.00 & 0.08 & 0.62 & 1.49 && 0.15 & 0.63 & 1.53 && 0.29 & 0.61 & 1.57 && 0.04 & 0.93 & 3.07 &-0.06 \\
   \hline
 \end{tabular}
\end{small}
\caption{\textit{$K=13$ studies}. Simulated comparisons between $T_1$, $T_3$, the fixed effects model estimator based on the $F$-distribution (FE) and the random effects model estimator (RE)  for bias, coverage probability (CP) and confidence interval width (W) for 1,000 simulated runs for various choices of $\rho$. The bias for the estimator of $\tau$ (Bias$_\tau$) is also included for the RE model.
The number of studies and sample sizes chosen are set equal to those for the comparison between the BB and Bb/bb groups from Table \ref{table:TH04}.}\label{table:BMD_sim}
\end{table}

In Table \ref{table:BMD_sim} we report the results for data simulated from $K=13$ studies where the sample sizes are chosen to be the same as those for comparison between the BB and Bb/bb groups from Table \ref{table:TH04}.  For simplicity we consider four estimators.  Firstly we consider the interval estimators for $\rho$ based on the $T_1$ and $T_3$ transformations (similar results were achieved for $T_2$ and $T_4$ and are therefore omitted) with weights according to \eqref{weights}.  For $T_1$, this is then also identical to the MLE estimators of $\rho$ based on the log transformation that is reported in Section \ref{section:MLE_norm}.  Given that our simulations have shown that $T_3$ can be a slightly better transformation function to achieve approximate normality, we have chosen this transformation for comparison to see if slightly better performance can be gained.  The other two methods are the MLE estimator based on the re-scaled $F$-distribution detailed in Section \ref{sect:MLE_F} and the MLE estimator based on the random effects model introduced in Section \ref{sect:REM}.  Throughout we will refer to these methods as $T_1$, $T_3$, $FE$ and $RE$ respectively.

For each of the methods highlighted above, we consider three choices for the variance parameter ($\tau^2$) for the random effects model.  The first choice, $\tau^2=0$, equates to the fixed effects model for which $T_1$, $T_2$ and $FE$ are well suited where $\tau$ is not estimated.  The other choices are for $\tau^2=0.2^2$ and $\tau^2=0.4^2$ which is a scenario favoring the RE approach.  The data is simulated according to the assumed data model (not the approximate normal model) in that the ratio of sample variances are sampled from the re-scaled $F$-distribution.

When $\tau=0$, all approaches achieve very close to nominal coverage of 0.95, small bias in estimating $\rho$ and similar interval widths.  The RE estimator of has an approximate bias of around 0.05 for $\tau$ which is somewhat expected given that the minimum for the estimate of $\tau$ is $0$.  The result is a slightly conservative interval for $\rho$.  In the presence of a random effect (i.e. when $\tau=0.2$ or $0.4$ in this simulation), we observe much lower than nominal coverages for $T_1$, $T_2$ and $FE$.  While the bias in estimating $\rho$ is often small, these estimators assume that $\tau=0$ which results in a smaller variance for the estimators and subsequently intervals that are too narrow.  The FE method performs typically worse with respect to the bias in estimator $\rho$, indicating that the transformations can adequately isolate the component $\rho$ from the missing random effect.  Close to nominal coverage is achieved when using the RE estimator although $\tau$ tends to be underestimated.

\setlength{\tabcolsep}{4pt}
\begin{table}[h!t]
\centering
\begin{small}
 \begin{tabular}{cccccccccccccccccc}
   \cmidrule{3-18}
&& \multicolumn{3}{c}{$T_1$}& & \multicolumn{3}{c}{$T_3$}& & \multicolumn{3}{c}{$FE$}& & \multicolumn{4}{c}{$RE$} \\
$\tau$&$\rho$&Bias&CP&W&&Bias&CP&W&&Bias&CP&W&&Bias&CP&W&Bias$_\tau$\\
\cmidrule{1-5} \cmidrule{7-9} \cmidrule{11-13} \cmidrule{15-18}
0.0& 0.20 &-0.00 & 0.94 & 0.04 &&-0.00 & 0.94 & 0.04 &&-0.00 & 0.94 & 0.04 &&-0.00 & 0.96 & 0.05 & 0.04 \\
   & 0.50 & 0.00 & 0.96 & 0.10 && 0.00 & 0.96 & 0.10 && 0.00 & 0.96 & 0.10 && 0.00 & 0.98 & 0.12 & 0.04 \\
   & 0.80 & 0.00 & 0.95 & 0.17 && 0.00 & 0.95 & 0.17 && 0.00 & 0.95 & 0.17 && 0.00 & 0.97 & 0.18 & 0.04 \\
   & 1.00 & 0.00 & 0.94 & 0.21 && 0.00 & 0.95 & 0.21 && 0.00 & 0.95 & 0.21 && 0.00 & 0.97 & 0.23 & 0.04 \\
   & 1.25 & 0.01 & 0.95 & 0.26 && 0.01 & 0.95 & 0.26 && 0.00 & 0.94 & 0.26 && 0.00 & 0.97 & 0.29 & 0.04 \\
   & 1.50 & 0.00 & 0.94 & 0.31 && 0.00 & 0.94 & 0.31 && 0.00 & 0.94 & 0.31 && 0.00 & 0.96 & 0.34 & 0.04 \\
   & 2.00 &-0.00 & 0.94 & 0.41 &&-0.00 & 0.95 & 0.42 &&-0.00 & 0.95 & 0.41 &&-0.00 & 0.96 & 0.46 & 0.04 \\
   & 5.00 & 0.01 & 0.95 & 1.04 && 0.01 & 0.95 & 1.04 && 0.01 & 0.95 & 1.04 && 0.02 & 0.97 & 1.15 & 0.04 \\   \cmidrule{1-5} \cmidrule{7-9} \cmidrule{11-13} \cmidrule{15-18}
0.2& 0.20 & 0.00 & 0.84 & 0.04 && 0.00 & 0.84 & 0.04 && 0.00 & 0.83 & 0.04 &&-0.00 & 0.94 & 0.06 &-0.03 \\
   & 0.50 & 0.00 & 0.83 & 0.10 && 0.00 & 0.84 & 0.10 && 0.01 & 0.83 & 0.10 && 0.00 & 0.93 & 0.14 &-0.04 \\
   & 0.80 &-0.00 & 0.83 & 0.17 && 0.00 & 0.84 & 0.17 && 0.01 & 0.82 & 0.17 &&-0.00 & 0.94 & 0.23 &-0.03 \\
   & 1.00 & 0.00 & 0.83 & 0.21 && 0.01 & 0.82 & 0.21 && 0.01 & 0.81 & 0.21 &&-0.00 & 0.94 & 0.29 &-0.03 \\
   & 1.25 & 0.01 & 0.83 & 0.26 && 0.01 & 0.83 & 0.26 && 0.02 & 0.83 & 0.26 && 0.01 & 0.94 & 0.36 &-0.03 \\
   & 1.50 &-0.00 & 0.83 & 0.31 && 0.01 & 0.84 & 0.31 && 0.02 & 0.84 & 0.32 &&-0.00 & 0.92 & 0.42 &-0.04 \\
   & 2.00 & 0.01 & 0.85 & 0.42 && 0.02 & 0.85 & 0.42 && 0.03 & 0.84 & 0.42 && 0.01 & 0.94 & 0.57 &-0.04 \\
   & 5.00 &-0.01 & 0.83 & 1.04 && 0.01 & 0.83 & 1.04 && 0.05 & 0.81 & 1.05 &&-0.01 & 0.92 & 1.42 &-0.04 \\   \cmidrule{1-5} \cmidrule{7-9} \cmidrule{11-13} \cmidrule{15-18}
0.4& 0.20 & 0.00 & 0.61 & 0.04 && 0.00 & 0.60 & 0.04 && 0.01 & 0.58 & 0.04 && 0.00 & 0.94 & 0.08 &-0.02 \\
   & 0.50 & 0.00 & 0.61 & 0.10 && 0.01 & 0.61 & 0.11 && 0.03 & 0.58 & 0.11 && 0.00 & 0.94 & 0.21 &-0.02 \\
   & 0.80 & 0.01 & 0.61 & 0.17 && 0.02 & 0.61 & 0.17 && 0.05 & 0.58 & 0.18 && 0.01 & 0.94 & 0.34 &-0.02 \\
   & 1.00 & 0.01 & 0.61 & 0.21 && 0.03 & 0.60 & 0.21 && 0.06 & 0.58 & 0.22 && 0.01 & 0.93 & 0.43 &-0.02 \\
   & 1.25 & 0.00 & 0.60 & 0.26 && 0.02 & 0.60 & 0.27 && 0.06 & 0.57 & 0.27 && 0.00 & 0.94 & 0.53 &-0.02 \\
   & 1.50 & 0.01 & 0.63 & 0.31 && 0.03 & 0.64 & 0.32 && 0.08 & 0.61 & 0.33 && 0.00 & 0.93 & 0.63 &-0.03 \\
   & 2.00 & 0.03 & 0.61 & 0.42 && 0.06 & 0.61 & 0.43 && 0.12 & 0.55 & 0.44 && 0.02 & 0.93 & 0.85 &-0.02 \\
   & 5.00 & 0.04 & 0.62 & 1.05 && 0.12 & 0.61 & 1.07 && 0.28 & 0.57 & 1.10 && 0.02 & 0.94 & 2.12 &-0.02 \\
   \hline
 \end{tabular}
\end{small}
\caption{\textit{$K=26$ studies}. Simulated comparisons between $T_1$, $T_3$, the fixed effects model estimator based on the $F$-distribution (FE) and the random effects model estimator (RE)  for bias, coverage probability (CP) and confidence interval width (W) for 1,000 simulated runs for various choices of $\rho$. The bias for the estimator of $\tau$ (Bias$_\tau$) is also included for the RE model.
The number of studies is twice that used in Table \ref{table:BMD_sim} where sample sizes have been repeated twice.}\label{table:BMD_sim2}
\end{table}
We now repeat the simulation from Table \ref{table:BMD_sim} but this time we assume twice as many studies (i.e. $K=26$) and report the results in Table \ref{table:BMD_sim2}.  To achieve this, the $K=13$ sample sizes were repeated.  For $\tau=0$ we observe excellent coverage for the interval estimators of $\tau$ and where the intervals are, on average, narrower.  When $\tau>0$, poor coverage is again observed for all methods except for the RE estimator where excellent coverage is obtained.  The bias in estimating $\tau$ has also decreased.

\setlength{\tabcolsep}{4pt}
\begin{table}[h!t]
\centering
\begin{small}
 \begin{tabular}{cccccccccccccccccc}
   \cmidrule{3-18}
&& \multicolumn{3}{c}{$T_1$}& & \multicolumn{3}{c}{$T_3$}& & \multicolumn{3}{c}{$FE$}& & \multicolumn{4}{c}{$RE$} \\
$\tau$&$\rho$&Bias&CP&W&&Bias&CP&W&&Bias&CP&W&&Bias&CP&W&Bias$_\tau$\\
\cmidrule{1-5} \cmidrule{7-9} \cmidrule{11-13} \cmidrule{15-18}
0.0& 0.20 &-0.00 & 0.96 & 0.02 &&-0.00 & 0.96 & 0.02 &&-0.00 & 0.97 & 0.02 &&-0.00 & 0.97 & 0.02 & 0.02 \\
   & 0.50 & 0.00 & 0.94 & 0.05 && 0.00 & 0.94 & 0.05 && 0.00 & 0.94 & 0.05 && 0.00 & 0.96 & 0.05 & 0.02 \\
   & 0.80 & 0.00 & 0.96 & 0.07 && 0.00 & 0.96 & 0.07 && 0.00 & 0.96 & 0.07 && 0.00 & 0.97 & 0.08 & 0.02 \\
   & 1.00 & 0.00 & 0.95 & 0.09 && 0.00 & 0.95 & 0.09 && 0.00 & 0.95 & 0.09 && 0.00 & 0.97 & 0.10 & 0.02 \\
   & 1.25 &-0.00 & 0.95 & 0.11 &&-0.00 & 0.95 & 0.11 &&-0.00 & 0.95 & 0.11 &&-0.00 & 0.97 & 0.13 & 0.02 \\
   & 1.50 & 0.00 & 0.95 & 0.14 && 0.00 & 0.95 & 0.14 && 0.00 & 0.95 & 0.14 && 0.00 & 0.96 & 0.15 & 0.02 \\
   & 2.00 &-0.00 & 0.95 & 0.18 &&-0.00 & 0.95 & 0.18 &&-0.00 & 0.95 & 0.18 &&-0.00 & 0.97 & 0.20 & 0.02 \\
   & 5.00 &-0.01 & 0.94 & 0.46 &&-0.01 & 0.94 & 0.46 &&-0.01 & 0.94 & 0.46 &&-0.01 & 0.96 & 0.50 & 0.02 \\  \cmidrule{1-5} \cmidrule{7-9} \cmidrule{11-13} \cmidrule{15-18}
0.2& 0.20 & 0.00 & 0.58 & 0.02 && 0.00 & 0.58 & 0.02 && 0.00 & 0.57 & 0.02 && 0.00 & 0.95 & 0.04 &-0.01 \\
   & 0.50 & 0.00 & 0.56 & 0.05 && 0.00 & 0.57 & 0.05 && 0.01 & 0.56 & 0.05 && 0.00 & 0.93 & 0.10 &-0.01 \\
   & 0.80 &-0.00 & 0.57 & 0.07 && 0.00 & 0.56 & 0.07 && 0.01 & 0.56 & 0.07 &&-0.00 & 0.95 & 0.16 &-0.01 \\
   & 1.00 & 0.00 & 0.56 & 0.09 && 0.00 & 0.55 & 0.09 && 0.01 & 0.55 & 0.09 && 0.00 & 0.93 & 0.20 &-0.01 \\
   & 1.25 &-0.00 & 0.55 & 0.11 && 0.00 & 0.55 & 0.12 && 0.01 & 0.55 & 0.12 &&-0.00 & 0.93 & 0.25 &-0.01 \\
   & 1.50 &-0.00 & 0.53 & 0.14 && 0.00 & 0.54 & 0.14 && 0.01 & 0.54 & 0.14 &&-0.00 & 0.93 & 0.30 &-0.01 \\
   & 2.00 & 0.00 & 0.58 & 0.18 && 0.01 & 0.58 & 0.18 && 0.02 & 0.56 & 0.18 && 0.00 & 0.93 & 0.39 &-0.01 \\
   & 5.00 & 0.02 & 0.57 & 0.46 && 0.04 & 0.57 & 0.46 && 0.08 & 0.56 & 0.46 && 0.01 & 0.91 & 0.98 &-0.02 \\  \cmidrule{1-5} \cmidrule{7-9} \cmidrule{11-13} \cmidrule{15-18}
0.4& 0.20 & 0.00 & 0.35 & 0.02 && 0.01 & 0.33 & 0.02 && 0.01 & 0.29 & 0.02 && 0.00 & 0.93 & 0.07 &-0.01 \\
   & 0.50 & 0.00 & 0.32 & 0.05 && 0.01 & 0.33 & 0.05 && 0.03 & 0.31 & 0.05 && 0.00 & 0.95 & 0.17 &-0.02 \\
   & 0.80 & 0.01 & 0.33 & 0.07 && 0.02 & 0.32 & 0.07 && 0.05 & 0.29 & 0.08 && 0.00 & 0.93 & 0.28 &-0.01 \\
   & 1.00 & 0.00 & 0.33 & 0.09 && 0.02 & 0.31 & 0.09 && 0.05 & 0.30 & 0.10 && 0.00 & 0.94 & 0.35 &-0.01 \\
   & 1.25 & 0.01 & 0.34 & 0.12 && 0.03 & 0.35 & 0.12 && 0.07 & 0.34 & 0.12 && 0.00 & 0.95 & 0.43 &-0.01 \\
   & 1.50 & 0.00 & 0.32 & 0.14 && 0.03 & 0.34 & 0.14 && 0.08 & 0.32 & 0.14 &&-0.00 & 0.96 & 0.52 &-0.01 \\
   & 2.00 & 0.01 & 0.33 & 0.18 && 0.04 & 0.32 & 0.19 && 0.10 & 0.31 & 0.19 && 0.01 & 0.93 & 0.69 &-0.02 \\
   & 5.00 & 0.03 & 0.33 & 0.46 && 0.11 & 0.34 & 0.47 && 0.21 & 0.34 & 0.48 && 0.01 & 0.96 & 1.74 &-0.01 \\
   \hline
 \end{tabular}
\end{small}
\caption{\textit{Large sample, $K=26$ studies}. Simulated comparisons between $T_1$, $T_3$, the fixed effects model estimator based on the $F$-distribution (FE) and the random effects model estimator (RE)  for bias, coverage probability (CP) and confidence interval width (W) for 1,000 simulated runs for various choices of $\rho$. The bias for the estimator of $\tau$ (Bias$_\tau$) is also included for the RE model.
The number of studies is twice that used in Table \ref{table:BMD_sim} where sample sizes have been repeated twice and have been doubled.}\label{table:BMD_sim3}
\end{table}
\setlength{\tabcolsep}{4pt}

Finally, in Table \ref{table:BMD_sim3} we repeat the simulation again for 26 studies, but this time we double all of the sample sizes.  We see the same patterns of coverage as before but with typically narrower intervals.  $T_1$ does a good job at estimating $\rho$ in the presence of a random effect with small bias reported although the coverage is poor.  The RE method has again achieved very good results and the bias in estimating $\tau$ has decreased.

\section{Bone mass density example continued}

We now illustrate how the theory and simulations above inform and affect an analysis of the bone mineral density data
 from Section \ref{section:mot} when comparing the data from the BB and Bb/bb groups from Table \ref{table:TH04}.  For simplicity we focus only on the Paulson transformation $T_3$.

\begin{figure}[h!t]
\centering
\includegraphics[scale=0.95]{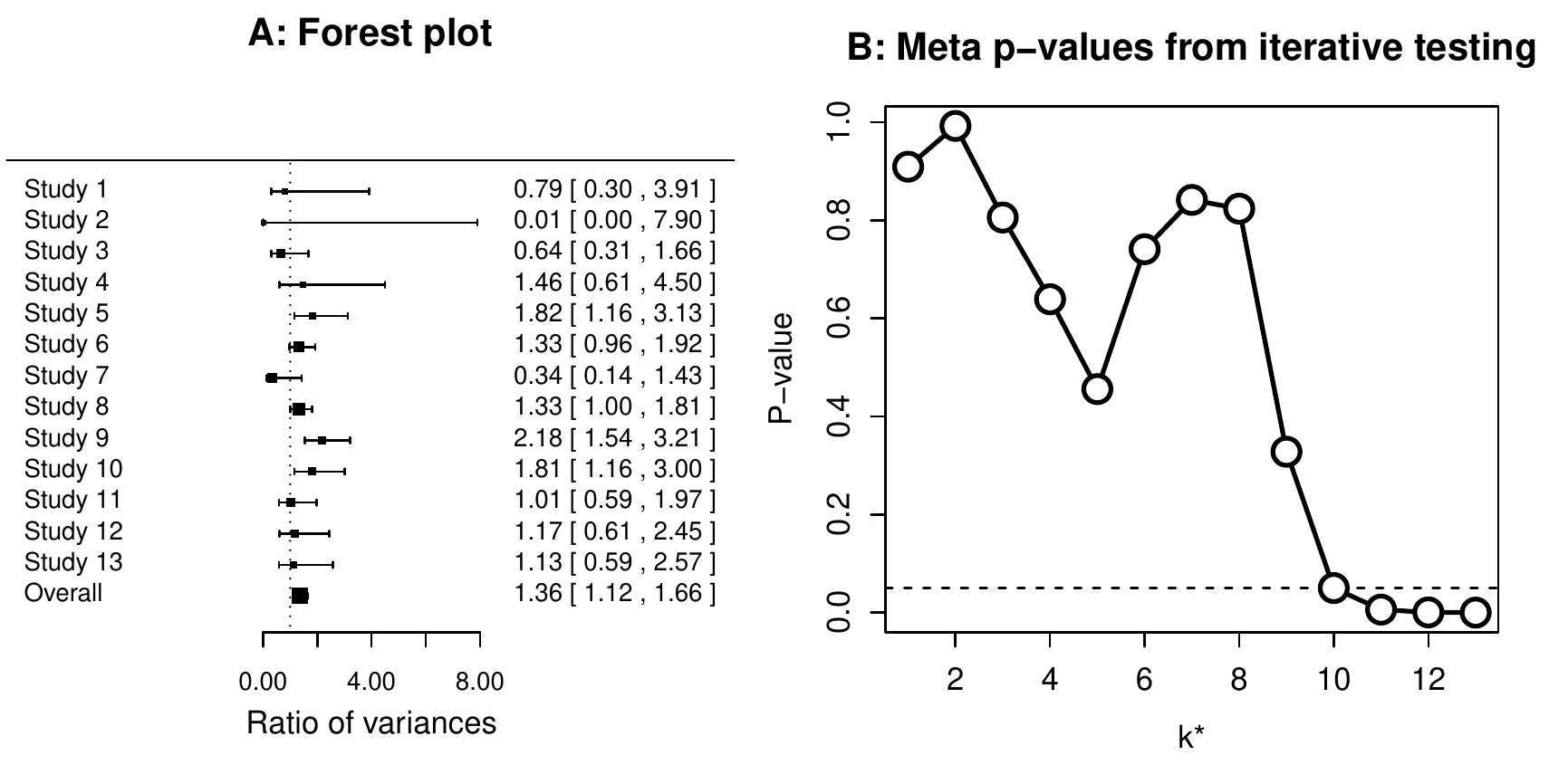}
\caption{Results of meta-analysis for the BB vs Bb/bb data in Table \ref{table:TH04}. Plot A depicts the forest plot for the analysis where the measure of interest is the ratio of variances.  The meta-analytic interval and estimate for a $\rho$ is given in the last row where the MLE estimator assuming the random effects model was used.  Plot B displays meta-analysis $p$-values for the test for equal variances when studies are added incrementally according to the magnitude of the $|Z_k|$s.}\label{figure4}\end{figure}

In Plot A of Figure \ref{figure4} we provide the forest plot for the meta analysis of the ratio of variances.  The study specific intervals are those arising from the standard $F$-test and the estimates are simply the ratio of estimated variances.  The estimate and interval labeled ``Overall" has arisen from the meta-analysis based on MLE estimation of the random effects model (see Section \ref{sect:REM}).  As can be seen from the forest plot, there appears to be a persistent suggestion that the true ratio is greater than one with two of the studies with the smaller estimates having high estimator variability and many of the studies with larger estimates having more precision. This is verified by the overall large estimate of 1.36 with corresponding 95\% confidence interval of $(1.12 , 1.66)$.  The estimate for the variance of the random effect is small $\widehat{\tau}^2=0.035$ with standard error 0.04.  When the MLE fixed effect model was assumed, the estimated $\rho$ was 1.45 with approximate 95\% confidence interval $(1.26, 1.68)$.  These results should lead to a rejection of using SMDs for the analysis given that the required assumption of equal variances in general is violated.  The forest plot is simple to create using the R package \texttt{metafor} \cite{VI10} since it includes functionality for the creation of forest plots that only requires the point and interval estimates.

As seen in the forest plot, it is possible that this highly significant result is due to the very large estimated ratio of variances from Study 9. In Plot B we explore this by iteratively adding studies based on the magnitude of their transformed statistics. The $p$-value in Plot B for $k^*=1$ is associated with the test based only on the value from Study 11 (the study with the smallest $Z_k$) and, not surprisingly, it is not significant.  For $k^*=2$, the test is run again but this time with both studies 11 and 1 (the two studies with the smallest in magnitude $Z_k$'s) and so on.  Here the weights used are calculated based only on the included studies.  There is a local minimum at $k^*=5$ before three negative $Z_k$'s are included which reduces the evidence against $H_0$.  These studies are associated with small weights when considering all $K=13$ studies.  However, when considering $k^*=8$ their weights are comparatively not so small allowing these studies to highly influence the test.  From here the larger studies are included and all with large positive $Z_k$'s.  Because of the sample sizes for these studies, they provide comparatively much more evidence for unequal variances.  We can see that there is very strong evidence for unequal variances even if Study 9 was ignored.  This means it would not simply suffice to remove Study 9 if SMDs were to be used since there is further evidence of a violation of equal variances.

\section{Discussion}

For a meta-analysis comparing means of two independent groups, a common approach is to consider SMDs which requires that the population variances within each of the groups within each study are equal.  However, when the population variances are not equal, the validity of the meta-analysis should be queried and the interpretation of the meta estimated effect may be unreliable.  In this paper we have shown that the meta-analysis of ratio of variances can be used to assess the validity of the equal variances assumption and consequently provide researchers with a justification to move towards other, perhaps less common, estimated effects.  Simple Q-Q plots of transformed ratios of sample variances allow visual
 inspection to determine departures from this assumption.  Alternatively, we also introduced meta-estimates of the ratio of variances in both the fixed effect and random effect settings.  Simulations show that the estimators exhibit very good  properties thus supporting their use in practice.

\appendix

\section{Technical details for the first two VSTs}

The technical details for VSTs $T_1$ and $T_2$ are provided here.  The mean and variance of $S\sim F_{\nu_1,\nu_2}$ are (see, for e.g., p.326 of \cite{JO&KO&BA95})
\begin{eqnarray}
\nonumber
  \e [S] &=& \frac {\nu _2}{\nu _2-2} \qquad \hbox{ for } \quad  \nu _2> 2 \\ \nonumber
  \var [S] &=&  \frac {2\nu _2^2(\nu_1+\nu _2-2)}{\nu _1(\nu _2-2)^2(\nu _2-4)} \qquad \hbox{ for } \quad \nu _2> 4~.
\end{eqnarray}
The variance of $S$ can be written as function of its mean $\var [S]=g(\e [S])$ in a number of ways, including:
\begin{eqnarray} \nonumber
  g_1(t) &=& c_1\,t^2 \qquad \hbox{where} \quad c_1=\frac {2(\nu_1+\nu _2-2)}{\nu _1(\nu _2-4)}, \\ \nonumber
  g_2(t) &=& c_2\,t\left (t+\frac{\nu _2}{\nu _1}\right )   \qquad \hbox{where} \quad c_2=\frac {2}{\nu _2-4}~.
\end{eqnarray}
This leads (see page 32 of \cite{BI&DO77}) to the respective VSTs $h_1(x)=\ln (x)/\sqrt {c_1}\, $ and
$h_2(x)=2\ln \{2(\sqrt x\,_+\sqrt {x+\nu _2/\nu _1}\,)\}/\sqrt {c_2}\, . $  Using the approximation given as $\e [h(S)]\doteq h(\e [S])+ h''(\e [S])\,\var [S]/2$, we obtain
\begin{eqnarray} \nonumber
 \e[h_1(S)] &=&   \frac {1}{\sqrt {c_1}\,}\;\ln (\e[S])- \frac {\var[S]}{2(\e[S])^2}=\frac {1}{\sqrt {c_1}\,}\;\ln (\e[S])-\frac {\sqrt {c_1}\,}{2}~.\\ \nonumber
  \e[h_2(S)] &=&   \frac {2}{\sqrt {c_2}\,}\;\ln \left \{2 \left (\sqrt {\e[S]}\,+
  \sqrt {\e[S]+\frac {\nu _2}{\nu _1}}\, \right )\right \} - \frac{\sqrt {c_2}\,(2\nu _1+\nu _2-2)}{4\sqrt {\nu _1^2+2\nu _1\nu _2-2\nu _1}\,}\nonumber
\end{eqnarray}
which leads to \eqref{T1} and \eqref{T2} after re-centering to achieve approximate mean zero.

\section{Variances for the random effects model MLE}\label{appendix:SE}

For simplicity when needed let $v=\tau^2$.  Using the notation from Section \ref{sect:REM},
\begin{align*}
\frac{\partial^2 l}{\partial \omega^2}=&-\sum^K_{k=1}\frac{1}{c_{1k}+\tau^2},\\
\frac{\partial^2 l}{\partial v^2}=&\sum^K_{k=1}\left[\frac{1}{2}\cdot\frac{1}{(c_{1k}+\tau^2)^2}-\frac{(y_k-\mu_k)^2}{(c_{1k}+\tau^2)^3}\right],\\
\frac{\partial^2 l}{\partial \omega \partial v}=&-\frac{1}{2}\sum^K_{k=1}\frac{y_k-\mu_k}{(c_{1k}+\tau^2)^2}.
\end{align*}

Taking the expectation with respect to random $\ln(S_1),\ldots,\ln(S_K)$ gives $E[\partial^2 l/(\partial v^2)]=-(1/2)\sum^{K}_{k=1}(c_{1k}+\tau^2)^{-1}$ and $E[\partial^2 l/(\partial \rho \partial v)]=0$.  By inverting the negative of the information matrix, we then have approximate variances for the MLEs.

\bibliographystyle{authordate4}
\bibliography{metaFrefs2}

\begin{thebibliography}{}

\bibitem[\protect\citename{Bickel \& Doksum, }1977]{BI&DO77}
{\sc Bickel, P~J, \& Doksum, K~A}. 1977.
\newblock {\em Mathematical statistics: Basic ideas and selected topics}.
\newblock San Francisco: Holden--Day.

\bibitem[\protect\citename{Borenstein {\em et~al.}, }2009]{BO&HE&HI&RO09}
{\sc Borenstein, M, Hedges, L~V, Higgins, J P~T, \& Rothstein, H~R}. 2009.
\newblock {\em Introduction to {M}eta-{A}nalysis}.
\newblock Chichester, West Sussex: Wiley.

\bibitem[\protect\citename{Cohen, }1988]{CO88}
{\sc Cohen, J}. 1988.
\newblock {\em {Statistical Power Analysis for the Behavioral Sciences (2nd
  Edition)}}. 2 edn.
\newblock Routledge.

\bibitem[\protect\citename{Hedges {\em et~al.}, }1999]{HE99}
{\sc Hedges, L~V, Gurevitch, J~A, \& Curtis, P~S}. 1999.
\newblock The meta-analysis of response ratios in experimental ecology.
\newblock {\em Ecology}, {\bf 80}(4), 1150--1156.
\newblock {DOI}: 10.2307/177062.

\bibitem[\protect\citename{Johnson {\em et~al.}, }1995]{JO&KO&BA95}
{\sc Johnson, NL, Kotz, S, \& Balakrishnan, N}. 1995.
\newblock {\em Continuous univariate distributions}.
\newblock  Vol. 2.
\newblock New York: John Wiley \& Sons.

\bibitem[\protect\citename{Kanis, }2002]{KA02}
{\sc Kanis, J~A}. 2002.
\newblock Diagnosis of osteoporosis and assessment of fracture risk.
\newblock {\em Lancet}, {\bf 359}, 1929--36.
\newblock {DOI}: 10.1016/S0140-6736(02)08761-5.

\bibitem[\protect\citename{Kulinskaya {\em et~al.}, }2008]{KMS-2008}
{\sc Kulinskaya, E, Morgenthaler, S, \& Staudte, R~G}. 2008.
\newblock {\em Meta analysis: a {G}uide to {C}alibrating and {C}ombining
  {S}tatistical {E}vidence}.
\newblock Wiley Series in Probability and Statistics.
\newblock Chichester: John Wiley \& Sons, Ltd.

\bibitem[\protect\citename{Kulinskaya {\em et~al.}, }2010]{KMS10}
{\sc Kulinskaya, E, Morgenthaler, S, \& Staudte, R~G}. 2010.
\newblock Variance stabilizing the difference of two binomial proportions.
\newblock {\em {American Statistician}}, {\bf 64}(4), 350--356.
\newblock {DOI}: 10.1198/tast.2010.09080.

\bibitem[\protect\citename{Malloy {\em et~al.}, }2011]{MA&PR&ST11}
{\sc Malloy, M~J, Prendergast, L~A, \& Staudte, R~G}. 2011.
\newblock Comparison of methods for fixed effect meta-regression of
  standardized differences of means.
\newblock {\em {Electronic Journal of Statistics}}, {\bf 5}, 83--101.
\newblock {DOI}: 10.1214/11-EJS598.

\bibitem[\protect\citename{Malloy {\em et~al.}, }2013]{MA&PR&ST13}
{\sc Malloy, M~J, Prendergast, L~A, \& Staudte, R~G}. 2013.
\newblock Transforming the model t: Random effects meta-analysis with stable
  weights.
\newblock {\em {Statistics in Medicine}}, {\bf 32}(11), 1842--1864.
\newblock {DOI}: 10.1002/sim.5666.

\bibitem[\protect\citename{Morgenthaler \& Staudte, }2012]{MO&ST12}
{\sc Morgenthaler, S, \& Staudte, R~G}. 2012.
\newblock Advantages of variance stabilization.
\newblock {\em {Scandinavian Journal of Statistics}}, {\bf 39}(4), 714--728.
\newblock {DOI}: 10.1111/j.1467-9469.2011.00768.x.

\bibitem[\protect\citename{Paulson, }1942]{PA42}
{\sc Paulson, F}. 1942.
\newblock An approximate normalization of the analysis of variance
  distribution.
\newblock {\em {Annals of Mathematical Statistics}}, {\bf 13}, 233--235.

\bibitem[\protect\citename{Prendergast \& Staudte, }2014]{prst-2014}
{\sc Prendergast, L~A, \& Staudte, R~G}. 2014.
\newblock Better than you think: interval estimators of the difference of
  binomial proportions.
\newblock {\em {Journal of Statistical Planning and Inference}}, {\bf 148},
  38--48.
\newblock {DOI}: 10.1016/j.jspi.2013.11.012.

\bibitem[\protect\citename{Thakkinstian {\em et~al.}, }2004]{TH04}
{\sc Thakkinstian, A, D'Este, C, Eisman, J, Nguyen, T, \& Attia, J}. 2004.
\newblock {Meta-Analysis of Molecular Association Studies: Vitamin D Receptor
  Gene Polymorphisms and BMD as a Case Study}.
\newblock {\em {Journal of Bone and Mineral Research}}, {\bf 19}, 419--428.

\bibitem[\protect\citename{Viechtbauer, }2010]{VI10}
{\sc Viechtbauer, W}. 2010.
\newblock Conducting meta-analyses in {R} with the {metafor} package.
\newblock {\em {Journal of Statistical Software}}, {\bf 36}(3), 1--48.
\newblock {DOI}: 10.1016/j.jspi.2013.11.012.

\bibitem[\protect\citename{Yuan \& Bentler, }2010]{YU&BE10}
{\sc Yuan, K~H, \& Bentler, P~M}. 2010.
\newblock Two simple approximations to the distributions of quadratic forms.
\newblock {\em {British Journal of Mathematical and Statistical Psychology}},
  {\bf 63}(2), 273--291.
\newblock {DOI}: 10.1348/000711009X449771.

\end{thebibliography}

\end{document}